\documentclass[a4paper,12pt]{article}

\usepackage{epsfig}
\usepackage{graphicx}

\usepackage{amsmath}
\usepackage{amssymb}

\begin{document}
\begin{titlepage}
\null
\begin{flushright}
HIP-2008-38/TH \\
December, 2008
\end{flushright}

\vskip 1.8cm
\begin{center}

 {\Large \bf Vortices, Q-balls and Domain Walls \\
\vskip 0.2cm
 on Dielectric M2-branes}

\vskip 2.3cm
\normalsize
\renewcommand\thefootnote{\alph{footnote}} 

  {\bf Masato Arai$^{\dagger}$\footnote{arai(at)sogang.ac.kr}, 
Claus Montonen$^\ddagger$\footnote{claus.montonen(at)helsinki.fi}
and Shin Sasaki$^\ddagger$\footnote{shin.sasaki(at)helsinki.fi}}

\vskip 0.5cm

  { \it 
  $^\dagger$Center for Quantum Spacetime (CQUeST), Sogang University, \\
  Shinsu-dong 1, Mapo-gu, Seoul 121-742, Korea\\
    \vskip 0.5cm
  $^\ddagger$Department of Physics,
  University of Helsinki \\
  Helsinki Institute of Physics,
  P.O.Box 64, FIN-00014, Finland \\
  }

\vskip 3cm

\begin{abstract}
We study BPS solitons in $\mathcal{N} = 6$ 
$U(N) \times U(N)$ Chern-Simons-matter theory deformed by an F-term mass. 
The F-term mass generically
 breaks $\mathcal{N} = 6$ supersymmetry down to $\mathcal{N} = 2$.
At vacua, M2-branes are polarized into a fuzzy $S^3$ 
forming a spherical M5-brane with topology $\mathbf{R}^{1,2} \times S^3$.
The polarization is interpreted as Myers' dielectric effect caused by 
an anti-self-dual 4-form flux $T_4$ in the eleven-dimensional supergravity.
Assuming a polarized M2-brane configuration, the model effectively reduces to the well-known 
abelian Chern-Simons-Higgs model studied in detail by Jackiw-Lee-Weinberg.
We find that the potential for the fuzzy $S^3$ radius agrees with 
the one calculated from the M5-brane point of view at large $N$.
This effective model admits not only BPS topological vortex and domain wall 
solutions but also non-topological solitons that keep 1/4 of the manifest 
$\mathcal{N} = 2$ supersymmetry. 
We also comment on the reduction of our configuration to ten dimensions.
\end{abstract}

\end{center}
\end{titlepage}

\newpage
\setcounter{footnote}{0}
\renewcommand\thefootnote{\arabic{footnote}} 
\section{Introduction}
No one doubts that understanding various aspects of low-energy effective 
theory of M2-branes enables us to uncover mysterious properties of 
M-theory and non-perturbative physics of string theories. 
Recently, Bagger, Lambert and Gustavsson proposed a three-dimensional $\mathcal{N} = 
8$ superconformal Chern-Simons model based a novel 3-algebra (BLG model) 
\cite{BaLa, 
Gu}. This model has been expected to describe the low energy effective 
theory of two coincident M2-branes in eleven dimensions. However, it seemed 
difficult to generalize their model to the case of arbitrary number of M2-branes.

Meanwhile, Aharony-Bergman-Jafferis-Maldacena proposed a 
three-dimensional $\mathcal{N} = 6$ $U(N) \times U(N)$ 
Chern-Simons-matter theory with level $(k,-k)$ (ABJM model) which 
may be regarded as the low-energy effective theory of $N$ coincident 
M2-branes probing $\mathbf{C}^4/\mathbf{Z}_k$ orbifold \cite{AhBeJaMa}. 
The model consists of gauge fields $A_{\mu}, \hat{A}_{\mu}$, four 
complex scalars $Z^A, W_A \ (A=1,2)$ in the (anti) bi-fundamental 
representation of the gauge group, and their superpartners. 
The model is believed to be a dual description of M-theory on $AdS_4 
\times S^7/\mathbf{Z}_k$.

After the proposal by Aharony-Bergman-Jafferis-Maldacena, a lot of works 
on the effective theory of interacting multiple membranes have been 
studied. Among them, classical solutions in 
the three-dimensional M2-brane world-volume theory attracted much attention 
since these solutions can be interpreted as various configurations of 
branes existing in the eleven-dimensional M-theory. 
In \cite{Te}, a BPS fuzzy funnel configuration that represents an 
M5-brane intersecting with multiple M2-branes was found\footnote{A similar analysis in the BLG model has been performed in \cite{KrMa}.}. 
A domain wall solution that interpolates between fuzzy $S^3$ and trivial 
vacua was found \cite{HaLi} in the mass deformed ABJM model that has maximal 
supersymmetry.
These solutions preserve half of $\mathcal{N} = 6$ supersymmetry.
On the other hand, time evolutions of fuzzy $S^3$ and M5/anti-M5 
configurations that are generically non-BPS were investigated in \cite{FuIwKoSa}.
These results provide some evidence that the ABJM model gives a correct 
description of the dynamics of multiple membranes. 

On the other hand, in \cite{Be}, $AdS_4 \times S^7$ dual theory of M2-branes with reduced 
supersymmetry was investigated following the prescription studied by Polchinski-Strassler \cite{PoSt}.
The author of \cite{Be} proposed a mass deformation of multiple membrane theory with 
$\mathcal{N} = 2$ world-volume supersymmetry and studied a probe 
M5-brane in the presence of anti-self-dual 4-form flux $T_4$ in the 
eleven-dimensional supergravity. 
The vacua are spherical M5-branes with topology $\mathbf{R}^{1,2} \times S^3$ 
sharing three dimensions with M2-branes. He conjectured 
that there exists BPS domain walls that interpolate between various 
vacua in the model. 
However, the correct description of the mass deformed multiple M2-brane theory 
was not known at that time.

In this paper, we study the ABJM model deformed by an F-term mass 
which was first introduced in \cite{GoRoRaVe}.
The F-term mass generically breaks $\mathcal{N} = 6$ supersymmetry down 
to $\mathcal{N} = 2$ preserving $SU(2)_{\mathrm{diag}}$ global symmetry of 
$SU(2) \times SU(2)$ and 
the resulting model is identified with the model studied in \cite{Be}.
We find that a vacuum configuration of the model is a fuzzy $S^3$.
We see that the radius of the sphere derived here coincides with the one 
obtained in \cite{Be} at large-$N$ identifying our vacuum as a spherical 
M5-brane with topology $\mathbf{R}^{1,2} \times S^3$. 
From the viewpoint of the M2-brane world-volume theory,
this is nothing but Myers' dielectric effect \cite{My} caused by 
eleven-dimensional supergravity flux $T_4 \sim m$. 
We also see that reducible configurations found in \cite{GoRoRaVe} are identified with a 
set of fuzzy $S^3$ shells.

In the latter half of this paper, 
we study various classes of BPS configurations with polarized 
M2-brane geometry in the mass deformed ABJM model.
Assuming a polarized M2-brane configuration, 
 we find that the Hamiltonian effectively reduces to that 
of the well-known abelian Chern-Simons-Higgs system with sixth-power 
potential studied in detail by Jackiw-Lee-Weinberg \cite{JaLeWe}. 
It has been known that this model exhibits an $\mathcal{N} = 2$ 
supersymmetry in three dimensions \cite{LeLeWe} and 
admits BPS topological vortices and domain walls.
In addition, there exists non-topological 
soliton solutions \cite{LePa} and a BPS supertube solution \cite{KiLeYe}.
We find that the potential for the $S^3$ radius in the effective Hamiltonian has 
the same structure as found in \cite{Be}. 
Since the configurations corresponding to the BPS solutions 
can exist only in the case of non-zero $m$, 
these solutions are supported against collapse by the non-zero background flux $T_4$ 
realizing a generalization of the Myers effect.

The organization of this paper is as follows. In section 2, we introduce the ABJM model deformed by the F-term mass.
Equations of motion and $\mathcal{N} = 2$ on-shell supersymmetry 
transformations are derived.
In section 3, we study the vacuum structure of the model. 
We compare the radius of the fuzzy $S^3$ with the one found in \cite{Be} 
at large-$N$ finding agreement between them.
In section 4, assuming an ansatz, we derive the 
effective Hamiltonian that reduces to the 
abelian Chern-Simons-Higgs model studied in \cite{JaLeWe, HoKiPa}. 
The non-abelian property of fields is totally encoded into the ``BPS matrices'' first constructed in \cite{GoRoRaVe}. 
We then perform the Bogomol'nyi completion in the effective 
Hamiltonian combining the kinetic 
and potential terms and derive the BPS equations.
The consistency between these BPS equations and the full equations of motion, 
as well as the number of preserved supersymmetries are discussed.
In section 5, parts of the exact and numerical results 
of the BPS solutions are discussed. Possible interpretations of these solutions in 
terms of eleven-dimensional M-theory are briefly discussed 
in this section.
Section 6 is devoted to conclusions and discussions.
In appendix A, the explicit form of the BPS matrices is presented. 
The $\mathcal{N} = 2$ superfield formulation of the ABJM 
model can be found in appendix B. 
The effective Lagrangian including fermion parts on the polarized 
M2-branes can be found in appendix C.

\section{The ABJM model}
The ABJM model \cite{AhBeJaMa} is a (2+1)-dimensional $\mathcal{N} = 6$ 
superconformal Chern-Simons-Higgs model of level $(k, -k)$ with $U(N) 
\times U(N)$ gauge symmetry.
The model is expected to describe the low energy world-volume theory of $N$ 
coincident M2-branes probing $\mathbf{C}^4/\mathbf{Z}_k$. 
We employ the notation and convention of \cite{BeKlKlSm} but with a different 
normalization of the $U(N)$ gauge generators $T^a$ such that $\mathrm{Tr} 
[T^a T^b] = \frac{1}{2} \delta^{ab}$.
The bosonic part of the massless ABJM action is 
\begin{eqnarray}
S = S_{\mathrm{kin}} + S_{\mathrm{CS}} + S_{\mathrm{pot}}, \label{boson-action}
\end{eqnarray}
where each term is given by 
\begin{eqnarray}
S_{\mathrm{kin}} &=& \int \! d^3 x \ 
\mathrm{Tr} 
\left[
- (D_{\mu} Z^A) (D^{\mu} Z^A)^{\dagger} 
- (D_{\mu} W_A) (D^{\mu} W_A)^{\dagger}
\right], \label{boson1} \\
S_{\mathrm{CS}} &=& 
\frac{k}{4 \pi} \int \! d^3 x \
\mathrm{Tr} \ \epsilon^{\mu \nu \lambda}
\left[
A_{\mu} \partial_{\nu} A_{\lambda} + \frac{2i}{3} A_{\mu} A_{\nu} A_{\lambda}
- \hat{A}_{\mu} \partial_{\nu} \hat{A}_{\lambda} - \frac{2i}{3} \hat{A}_{\mu} \hat{A}_{\nu} \hat{A}_{\lambda}
\right], \label{boson2}
\end{eqnarray}
\begin{eqnarray}
S_{\mathrm{pot}} &=& - \frac{4\pi^2}{k^2} \int \! d^3 x \ 
\mathrm{Tr}
\left[
(Z^A Z^{\dagger}_A + W^{\dagger A} W_A) (Z^B Z^{\dagger}_B - W^{\dagger 
B} W_B) (Z^C Z^{\dagger}_C - W^{\dagger C} W_{C}) \right. \nonumber \\
& & \qquad \qquad \qquad \quad + (Z^{\dagger}_A Z^A + W_A W^{\dagger A}) (Z^{\dagger}_B Z^B - W_B 
W^{\dagger B}) (Z^{\dagger}_C Z^C - W_C W^{\dagger C}) \nonumber \\
& & \qquad \qquad \qquad \quad - 2 Z^{\dagger}_A (Z^B Z^{\dagger}_B - W^{\dagger B} W_B) Z^A 
(Z^{\dagger}_C Z^C - W_C W^{\dagger C}) \nonumber \\
& & \left.
\qquad \qquad \qquad \quad - 2 W^{\dagger A} (Z^{\dagger}_B Z^B - W_B W^{\dagger B}) W_A 
(Z^C Z^{\dagger}_C - W^{\dagger C} W_C)
\right] \nonumber \\
& & + \frac{16\pi^2}{k^2} \int \! d^3 x \ 
\mathrm{Tr}
\left[
W^{\dagger A} Z^{\dagger}_B W^{\dagger C} W_A Z^B W_C - W^{\dagger A} 
Z^{\dagger}_B W^{\dagger C} W_C Z^B W_A \right. \nonumber \\
& & \qquad \qquad \qquad \quad \left. + Z^{\dagger}_A W^{\dagger B} Z^{\dagger}_C Z^A W_B Z^C
- Z^{\dagger}_A W^{\dagger B} Z^{\dagger}_C Z^C W_B Z^A
\right]. \label{boson3}
\end{eqnarray}
Here $A_{\mu}$, $\hat{A}_{\mu}$ are $U(N) \times U(N)$ gauge fields, 
$Z^A, W^{\dagger A} \ (A=1,2)$ are complex scalar fields in the $U(N) \times U(N)$ bi-fundamental $(\mathbf{N}, 
\bar{\mathbf{N}})$ representation. 
The world-volume metric is $\eta_{\mu \nu} = \mathrm{diag} (-1,1,1)$.
The gauge covariant derivative is 
\begin{eqnarray}
& & D_{\mu} Z^A = \partial_{\mu} Z^A + i A_{\mu} Z^A - i Z^A 
\hat{A}_{\mu}.
\end{eqnarray}
The gauge field strength is defined as 
\begin{eqnarray}
F_{\mu \nu} = \partial_{\mu} A_{\nu} - \partial_{\nu} A_{\mu}
+ i [A_{\mu}, A_{\nu}],
\end{eqnarray}
and similarly for $\hat{A}_{\mu}$.
The common $U(1)$ charge is fixed to $+1$.
The model exhibits a manifest $SU(2) \times SU(2) \times U(1)_R$ global 
symmetry. Under the each $SU(2)$s, 
$Z^A, W_A$ transform independently in the 
fundamental representation.
Apart from this manifest symmetry, there is an $SU(2)_R$ symmetry 
under which the fields $Z^1, W^{\dagger 1}$ (and $Z^2, W^{\dagger 2}$) 
transform as a doublet. 
It was discussed in \cite{AhBeJaMa} that the 
$SU(2) \times SU(2)$ 
global symmetry 
is combined with the $SU(2)_R$ and enhanced to 
$SU(4)_R \sim SO(6)_R$. 
Therefore, 
for $k>2$, the model has $\mathcal{N} = 6$ supersymmetry.
We consider a trivial embedding of the world-volume in the space-time.
Namely, the world-volume coordinates $(x^0, x^1, x^2)$ are identified 
with the space-time coordinates $(X^0, X^1, X^2)$.
The four complex scalars $Z^A, W^{\dagger A}$ represent the transverse 
displacement of the M2-branes along eight directions $X^I \ (I = 3, 
\cdots, 10)$. The orbifolding symmetry $\mathbf{Z}_k$ act as 
$(Z^A, W^{\dagger A}) \to e^{\frac{2\pi i}{k}} (Z^A, W^{\dagger A})$.
The $\mathcal{N} = 2$ superfield formalism of the model can be 
found in Appendix B.

The Gauss' law constraint comes from the equation of motion for the 
gauge fields,
\begin{eqnarray}
& & \frac{k}{4\pi} \epsilon^{\rho \mu \nu} F_{\mu \nu} 
= i \left[
Z^A (D^{\rho} Z^A)^{\dagger} - (D^{\rho} Z^A) Z^{\dagger}_A
\right]
+ i 
\left[
W^{\dagger A} (D^{\rho} W_A) - (D^{\rho} W_A)^{\dagger} W_A
\right], \label{gauss1} \\
& & \frac{k}{4\pi} \epsilon^{\rho \mu \nu} \hat{F}_{\mu \nu} 
= i \left[
(D^{\rho} Z^A)^{\dagger} Z^A - Z^{\dagger}_A (D^{\rho} Z^A )
\right]
+ i 
\left[
(D^{\rho} W_A) W^{\dagger A} - W_A (D^{\rho} W_A)^{\dagger}
\right]. \label{gauss2}
\end{eqnarray}
The Noether current and charge corresponding to the 
following 
$U(1)$ gauge 
transformation (which we call baryonic $U(1)$)
\begin{eqnarray}
(Z^A, W^{\dagger A}) \longrightarrow e^{i \alpha} (Z^A, W^{\dagger A})
\label{baryonic_u1}
\end{eqnarray}
are derived as 
\begin{eqnarray}
& & j^{\mu}_b = - i \mathrm{Tr} 
\left[
Z^A D^{\mu} Z^{\dagger}_A - Z^{\dagger}_A D^{\mu} Z^A 
- W_A D^{\mu} W^{\dagger A} + W^{\dagger A} D^{\mu} W_A
\right], \\
& & Q_b = i \int \! d^2 x \ \mathrm{Tr} 
\left[
Z^A D_{0} Z^{\dagger}_A - Z^{\dagger}_A D_{0} Z^A 
- W_A D_{0} W^{\dagger A} + W^{\dagger A} D_{0} W_A
\right]. 
\label{baryonic_charge}
\end{eqnarray}

Let us consider massive deformations of the model. 
There are two kinds of massive deformations: 
F- and D-term deformations \cite{GoRoRaVe}\footnote{See also \cite{HoLeLeLePa} for massive 
deformations with reduced supersymmetries.}. 
Here we focus on the F-term mass deformation.
The superpotential is given by 
\begin{eqnarray}
W = W_0 + \delta W, 
\label{superpotential}
\end{eqnarray}
where
\begin{eqnarray}
& & W_0 = \frac{1}{4} \left(\frac{8\pi}{k} \right) 
\epsilon_{AC} \epsilon^{BD} 
\mathrm{Tr} \left[
\mathcal{Z}^A \mathcal{W}_B \mathcal{Z}^C \mathcal{W}_D
\right], \label{massless_superpot}
\end{eqnarray}
is the original superpotential in the massless ABJM model \cite{BeKlKlSm} and
\footnote{Note that its ``conjugate'' part is given by 
$\delta \bar{W} = - m \mathrm{Tr} [\bar{\mathcal{W}}^{A} \bar{\mathcal{Z}}_A]$.
The minus sign in front of $m$ comes from the extra minus sign of the 
components in a ``conjugated'' superfield \cite{BeKlKlSm}. See Appendix B.}
\begin{eqnarray}
\delta W = 
m \mathrm{Tr} 
\left[
\mathcal{Z}^A \mathcal{W}_A
\right], \quad m \in \mathbf{R},
\label{mass_superpot}
\end{eqnarray}
is a mass term which breaks the $SU(2) \times SU(2)$ global symmetry 
down to $SU(2)_{\mathrm{diag}}$ and keeps 
manifest the $\mathcal{N} = 2$ supersymmetry.
This is different from the D-term mass deformation which keeps 
the manifest $SU(2) \times SU(2) \times U(1)_R \times \mathbf{Z}_2$ 
symmetry and preserves $\mathcal{N} = 6$ maximal supersymmetry \cite{GoRoRaVe}.
Here the $\mathbf{Z}_2$ transformation swaps $Z^A$ and $W^{\dagger A}$.
It is easy to derive the equation of motion for the scalar fields. 
The full equation of motion for $Z^A$ is 
\begin{eqnarray}
&&D_{\mu} D^{\mu} Z^A = \left(
\frac{4\pi^2}{k^2}
\right)
\left[
(Z^B Z^{\dagger}_B - W^{\dagger B} W_B)^2 Z^A 
\right. \nonumber \\
& &~
+ (Z^C Z^{\dagger}_C - W^{\dagger C} W_C) (Z^B Z^{\dagger}_B 
 + W^{\dagger B} W_B) Z^A 
 + (Z^C Z^{\dagger}_C + W^{\dagger C} W_C)
 (Z^B Z^{\dagger}_B - W^{\dagger B} W_B) Z^A 
 \nonumber \\
& &~
+ Z^A (Z^{\dagger}_B Z^B - W_B W^{\dagger B})^2 
 + Z^A (Z^{\dagger}_B Z^B - W_B W^{\dagger B}) 
 (Z^{\dagger}_C Z^C + W_C W^{\dagger C}) \nonumber \\
& &~
+ Z^A (Z^{\dagger}_B Z^B + W_B W^{\dagger B}) 
 (Z^{\dagger}_C Z^C - W_C W^{\dagger C}) \nonumber \\
& &~
- 2 (Z^B Z^{\dagger}_B - W^{\dagger B} W_B) Z^A (Z^{\dagger}_C Z^C - 
W_C W^{\dagger C}) - 2 Z^B (Z^{\dagger}_C Z^C - W_C W^{\dagger C}) 
Z^{\dagger}_B Z^A \nonumber \\
& &~ 
- 2 Z^A Z^{\dagger}_C (Z^B Z^{\dagger}_B - W^{\dagger B} W_B) Z^C 
- 2 Z^A W_B (Z^C Z^{\dagger}_C - W^{\dagger C} W_C) W^{\dagger B} 
\nonumber \\
& &~ 
- 2 W^{\dagger C} (Z^{\dagger}_B Z^B - W_B W^{\dagger B}) W_C Z^A 
\nonumber \\
& &~ 
- 4 W^{\dagger C} W_B Z^A W_C W^{\dagger B} 
+ 4 W^{\dagger C} W_C Z^A W_B W^{\dagger B} 
- 4 W^{\dagger B} Z^{\dagger}_C Z^A W_B Z^C \nonumber \\
& &~ 
\left.
- 4 Z^C W_B Z^A Z^{\dagger}_C W^{\dagger B}
+ 4 W^{\dagger B} Z^{\dagger}_C Z^C W_B Z^A 
+ 4 Z^A W_B Z^C Z^{\dagger}_C W^{\dagger B}
\right] \nonumber \\
& &~ 
+ m^2 Z^A - \frac{4\pi m}{k} \epsilon^{AC} \epsilon_{BD} 
\left(
W^{\dagger B} W_C W^{\dagger D} 
+ Z^B Z^{\dagger}_C W^{\dagger D} 
+ W^{\dagger B} Z^{\dagger}_C Z^D
+ Z^B W_C Z^D
\right). \label{eom}
\end{eqnarray}
The equation of motion for $W^{\dagger A}$ is obtained by replacing $Z^A$ with 
$W^{\dagger A}$ in the above expression.

For later convenience, we derive the Hamiltonian of the model. 
The Chern-Simons part gives a vanishing contribution because it is a 
topological quantity. However, the gauge fields enter in the Hamiltonian 
through the covariant derivative. The result is
\begin{eqnarray}
H = \int d^3 x \ \mathrm{Tr}
\left[ |D_0 Z^A|^2 + |D_i Z^A|^2 
+ |D_0 W_A|^2 + |D_i W_A|^2
 + V_{\mathrm{scalar}}
\right],
\label{ABJM_hamiltonian}
\end{eqnarray}
where 
the potential part is
\begin{eqnarray}
V_{\mathrm{scalar}} &=& V_D + V_F, \\
V_D &=& 
\frac{4 \pi^2}{k^2} \mathrm{Tr} 
\left[
\left|
 Z^B Z^{\dagger}_B Z^A - Z^A Z^{\dagger}_B Z^B
- W^{\dagger B} W_B Z^A + Z^A W_B W^{\dagger B} 
\right|^2 
  \right. \nonumber \\
& & \qquad \qquad + \left.
\left|
W^{\dagger B} W_B W^{\dagger A} 
- W^{\dagger A} W_B W^{\dagger B} - Z^B Z^{\dagger}_B W^{\dagger A} 
+ W^{\dagger A} Z^{\dagger}_B Z^B
\right|^2
\right], \\
V_F &=& \frac{64 \pi^2}{k^2} \mathrm{Tr}
\left[
\left|
\frac{km}{8\pi} W_A + \frac{1}{2} 
\epsilon_{AC} \epsilon^{BD} W_B Z^C W_D
\right|^2 
+
\left|
\frac{km}{8\pi} Z^A - \frac{1}{2}
\epsilon^{AC} \epsilon_{BD} Z^B W_C Z^D
\right|^2
\right]. \nonumber \\
\end{eqnarray}

As we mentioned, the model exhibits at least $\mathcal{N} = 2$ manifest 
supersymmetry. It can be shown that the model is invariant under the 
following (on-shell) $\mathcal{N} = 2$ supersymmetry,
\begin{eqnarray}
& & \delta Z^A =\sqrt{2}\epsilon \zeta^A\,, \\
& & \delta W_A =\sqrt{2}\epsilon \omega_A, \\
& & \delta \zeta^A =
 \sqrt{2}\epsilon F^A+i\sqrt{2}\bar{\epsilon}(-Z^A\hat{\sigma}+\sigma
 Z^A)+\sqrt{2}i\gamma^\mu \bar{\epsilon}D_\mu Z^A\,, \label{susy-trans1}\\
& & \delta \omega_A =\sqrt{2}\epsilon G_A
 +i\sqrt{2}\bar{\epsilon}(-W_A\sigma+\hat{\sigma}W_A)
 +\sqrt{2}i\gamma^\mu \bar{\epsilon}D_\mu W_A\,, \label{susy-trans2}\\
& & \delta A_{\mu} =-{i \over \sqrt{2}}(\epsilon \gamma_\mu
 \bar{\chi})-{i \over \sqrt{2}}(\bar{\epsilon}\gamma_\mu \chi)\,, \\
& & \delta \hat{A}_{\mu} =-{i \over \sqrt{2}}(\epsilon \gamma_\mu
 \hat{\bar{\chi}})-{i \over \sqrt{2}}(\bar{\epsilon}\gamma_\mu \hat{\chi})\,,
\end{eqnarray}
where 
$\zeta^A, \omega_A$ are fermionic partners, and 
$F^A, G_A, \sigma^a, \hat{\sigma}^a, \chi, \hat{\chi}$ are auxiliary fields whose equations of motion are given in
 (\ref{aux1})-(\ref{aux5}) in appendix B. 
The supersymmetric transformation is achieved by the operator 
$\delta = \epsilon Q + \bar{\epsilon} \bar{Q}$ where $Q_{\alpha}$ is a  
supercharge and $\epsilon$ is a two component complex spinor.

\section{Vacua as dielectric M2-branes}
In this section, we discuss the physical meaning of our F-term deformation 
and investigate the vacuum structure of the model.
For the massless ABJM model, there is a dual description as M-theory on 
$AdS_4 \times S^7$ at large-$N$ \footnote{We here focus on the $k=1$ sector where 
the orbifolding by $\mathbf{Z}_k$ becomes trivial. However, we expect 
that the results in this section hold even for the $k> 1$ case.}.
We can perturb the $AdS_4 \times S^7$ side by turning on a non-trivial 
supergravity flux which causes additional terms on the dual M2-brane side. 
Indeed, in \cite{Be}, Bena considered a mass 
perturbation of the theory of $N$ coincident M2-branes.
Its fermionic part is given by
\begin{eqnarray}
\delta \mathcal{L} = \mathrm{Re} 
(
m \sum_{I=1}^4 \Lambda_I^2
). \label{fermion_mass}
\end{eqnarray}
Here $\Lambda_I$ are complexified Majorana fermions living on the M2-branes.
The mass term keeps 
at least $\mathcal{N} = 2$ supersymmetry of the $\mathcal{N} = 8$ maximal supersymmetry. 
The mass terms for the scalar fields break $SO(8)_R$ symmetry down to $SO(4)_R$.
We will discuss this issue of global symmetry a little bit in detail later.
The author of \cite{Be} conjectured that the deformed model is given by a superpotential of 
the form 
\begin{eqnarray}
W \sim z_1 z_2 z_3 z_4 + m \sum_{I=1}^4 z_I^2, \label{csuperpot}
\end{eqnarray}
where $z_a = x_{a+2} + i x_{a+6}, \ (a=1, \cdots, 4)$ are $\mathcal{N} = 
2$ chiral superfields whose lowest components are 
complex scalars representing fluctuations along transverse directions to the M2-brane 
world-volume. 
The non-zero mass parameter on the 
M2-brane side was identified with non-trivial anti-self-dual constant 4-form flux $T_4$
on the supergravity side via $AdS$/CFT duality. 
The 4-form flux has a non-trivial value in the directions transverse 
to the M2-branes.
The ``(anti) self-dual'' means that the flux satisfies
\begin{eqnarray}
\frac{1}{4!} \epsilon_{ijkl} {}^{mnop} T_{mnop} = \pm T_{ijkl},
\end{eqnarray}
where $\epsilon_{ijkl} {}^{mnop}$ is the eight-dimensional epsilon 
tensor and indices $i, j, \cdots $ run from 3 to 10.

One may imagine that the M2-branes are polarized due to the 
background flux forming another higher dimensional brane in eleven-dimensions.
The most natural candidate of this higher dimensional brane is an M5-brane. Indeed, 
the vacuum structure of this deformed model was studied through an 
examination of a probe M5-brane with $N$-units of M2-brane charge in the 
$AdS_4 \times S^7$ perturbed by the flux. 
For the case of four equal masses, which is our concern here, the 
supersymmetric vacuum is a configuration of an M5-brane with 
geometry $\mathbf{R}^{1,2} \times S^3$.
Its $\mathbf{R}^{1,2}$ directions are shared with the M2-branes while 
other three-dimensions are wrapped on an $S^3$ inside the $S^7$.
The radius $r$ of the $S^3$ was derived by evaluating the M5-brane action 
via the Pasti-Sorokin-Tonin and Perry-Schwarz approaches \cite{PaSoTo, PeSc}. 
Its ``potential'' term at large-$N$ is \cite{Be}
\begin{eqnarray}
V (z) = \frac{3 T_5}{4 A} (r^3 - 4 r m A/3)^2, \label{pot-M5}
\label{radial_potential}
\end{eqnarray}
where $A = 4 \pi N/M_{11}^3$ and $M_{11}$ is the eleven-dimensional Planck mass.
Here we have added the overall M5-brane tension $T_5$, which was 
ignored in \cite{Be}, 
and we have restricted ourselves to the 3456 plane in the
 3-7, 4-8, 5-9, 6-10 planes using $SO(4)$ rotations \cite{Be}.
From the result (\ref{radial_potential}), 
we find a supersymmetric vacuum
\begin{eqnarray}
r^2 \sim N \frac{m}{M_{11}^3}. 
\label{M5_radius}
\end{eqnarray}
Apart from the non-zero radius, 
there is also a trivial vacuum $r = 0$. Therefore there should exist a 
domain wall solution which interpolates between these vacua.
It was conjectured that the tension of a domain wall is
\begin{eqnarray}
\tau_{DW} \sim m^2 N^2. \label{dw_tension}
\end{eqnarray}
The above situation can be extended to the case that the $N$ M2-branes
are uniformly distributed on more than one 3-sphere, with M5-brane charges.
The potential felt
 by a probe M5-brane with M2-brane charges is still written by
 (\ref{pot-M5}), but with a different radius $r_b$ and M2-brane charges
 $n_b$.
Its form is given by (once again restricting to the 3456 plane),
\begin{eqnarray}
 V=\sum_b {3T_5 \over 4A_b}(r_b^3-4r_bmA_b/3)^2,\label{pot-M2-shell}
\end{eqnarray}
where $A_b\equiv 4\pi n_b/M_{11}^3$.
Therefore the ground states form M2-brane 3-sphere shells with 
 radius $r_b=4mA_b/3$.

With these results in mind, let us return to our model.
The ABJM model deformed by the F-term mass is given by the 
superpotential (\ref{superpotential}). 
This is just the superpotential (\ref{csuperpot}) conjectured in \cite{Be}. 
Its fermionic part 
$\mathrm{Re} ( 2 m \mathrm{Tr} [ \zeta^A \omega_A])$ 
is nothing but the equation (\ref{fermion_mass}) via the 
identification 
\begin{eqnarray}
\zeta^A = \frac{1}{\sqrt{2}} (\Lambda^A + i \Lambda^{A+2}), \quad 
\omega_A = \frac{1}{\sqrt{2}} (\Lambda^{A} - i \Lambda^{A+2}).
\end{eqnarray}
The mass perturbation part (\ref{mass_superpot}) 
keeps $SU(2)_{\mathrm{diag}}$ which is a subgroup of 
the expected R-symmetry group $SO(4)_R$. 
However, we remind the fact that for the massless ABJM model, 
the global $SU(2) \times SU(2)$ symmetry does not commute with the 
$SU(2)_R$, they combine giving an $SU(4)_R$ symmetry. 
This mechanism of symmetry enhancement 
would be true even for our case. 
Although it is not manifest here, 
the remaining $SU(2)_{\mathrm{diag}}$ 
global symmetry would be combined with $SU(2)_R$ generating 
$SU(2)_{\mathrm{diag}} \times SU(2)_R \sim SO(4)_R$. 

The supersymmetric vacuum of our model can be obtained as follows \cite{GoRoRaVe}.
Assuming a configuration
\begin{eqnarray}
Z^A = W^{\dagger A} = f_0 S^A, 
\ f_0 \in \mathbf{C},
\label{vac_ansatz}
\end{eqnarray}
the condition $V_D = 0$ is automatically satisfied. 
Here $S^A$ are BPS matrices \cite{GoRoRaVe} presented in appendix A.
The F-term condition
\begin{eqnarray}
 \frac{\partial \mathcal{W}}{\partial Z^A} 
    =- \left( \frac{k m}{8 \pi} - \frac{1}{2} |f_0|^2 \right) \bar{f}_0 
    S^{\dagger}_A=0,\quad
 \frac{\partial \mathcal{W}}{\partial W_A}=
 - \left( \frac{k m}{8 \pi} - \frac{1}{2} |f_0|^2 \right) f_0 S^A=0
\end{eqnarray}
yields the following vacua 
\begin{eqnarray}
A_{\mu} = \hat{A}_{\mu} = 0, 
\quad 
|f_0|^2 = 0, \ \frac{km}{4 \pi}, \label{vacuum}
\end{eqnarray}
where $f_0 = 0$ is a trivial vacuum. 
At the vacuum corresponding to $f_0 \not= 0$, 
the $U(N) \times U(N)$ gauge symmetry is broken to $U(1) \times U(1)$.
One of these $U(1)$s is the ``baryonic $U(1)$'', eq. 
(\ref{baryonic_u1}), and the other one being an independent rotation of 
a degree of complexity freedom which decouples when the ansatz (\ref{vac_ansatz}) is used.
To see the physical meaning of $f_0 
\not= 0$ solution, let us define the following combinations of the transverse 
fluctuation modes
\begin{eqnarray}
U^A_{\pm} \equiv \frac{1}{2} (Z^A \pm W^{\dagger A}), \ (A = 1, 2).
\end{eqnarray}
Due to the ansatz $Z^A = W^{\dagger A}$, only $U^A_{+}$ is non-zero. 
Therefore 
half of the eight transverse degrees of freedom 
$X^I \ (I = 3, \cdots, 10)$ 
is now dropped and 
this configuration is nothing but a fuzzy $S^3$ with radius
\begin{eqnarray}
R^2 = \frac{2}{N T_2} \mathrm{Tr} [U_{+}^A U^{\dagger}_{+A}] = \frac{1}{T_2} (N-1) \frac{km}{2\pi} \sim N \frac{m}{M_{11}^3} \quad (N \to \infty).
\end{eqnarray}
Here $T_2$ is a tension of an M2-brane. 
We have used the relation that the $p$-brane tension in 
eleven-dimensions is given by $T_p = (2\pi)^{-p} M_{11}^{p+1}$.
This result coincides with the equation (\ref{M5_radius}) 
obtained from the dual M5-brane description.
Therefore the vacuum configuration (\ref{vacuum}) is interpreted as  
polarized M2-branes caused by the background flux $T_4$ with its VEV $m$. 
This is just an M-theory realization of Myers' dielectric effect 
\cite{My}. 

Because the model admits vacua $f_0 = 0, \ f_0 \not= 0$
and in each sector they have different values of the superpotential, there is a domain wall solution 
that interpolates between them.
Indeed, it is easy to find such a solution with vanishing gauge fields.
The Bogomol'nyi completion of the energy density for the configuration 
$Z^A = Z^A (x_1), \ W_A = W_A (x_1)$ with vanishing gauge fields is
\begin{eqnarray}
E &=& \int \! d x^1 \ \mathrm{Tr}
\left[
\partial_1 Z^A \partial_1 Z^{\dagger}_A + \partial_1 W_A \partial_1 
W^{\dagger A} + V_D + V_F
\right] 
\nonumber \\
&=& \int \! d x^1 \ \mathrm{Tr} 
\left[
\left|
\partial_1 Z^A - m W^{\dagger A} + \frac{4\pi}{k} 
\epsilon^{AC} \epsilon_{BD} W^{\dagger B} Z^{\dagger}_C W^{\dagger D}
\right|^2 \right. \nonumber \\
& & \qquad \qquad 
\left.  + 
\left|
\partial_1 W^{\dagger A} - m Z^A + \frac{4\pi}{k} \epsilon^{AB} 
\epsilon_{CD} Z^C W_B Z^D
\right|^2 + V_D \right] + T,
\label{domain_wall_completion}
\end{eqnarray}
where the surface term is
\begin{eqnarray}
T &=& - \int \! d x^1 \ \partial_1 
\mathrm{Tr} 
\left[
- m Z^A W_A - \frac{2\pi}{k} \epsilon_{AC} \epsilon^{BD} Z^A W_B Z^C W_D 
\right. \nonumber \\
& & \qquad \qquad \qquad \qquad \left.
- m W^{\dagger A} Z^{\dagger}_A + \frac{2\pi}{k} \epsilon^{AC} \epsilon_{BD}
Z^{\dagger}_A W^{\dagger B} Z^{\dagger}_C W^{\dagger D}
\right].
\end{eqnarray}
The BPS equations are obtained by requiring that 
$V_D = 0$ and the vanishing condition 
inside the absolute values in the equation (\ref{domain_wall_completion}).
A solution to these equations is given by 
\begin{eqnarray}
Z^A = W^{\dagger A} = f (x_1) S^A, 
\quad 
f^2 (x_1) = \frac{k}{4\pi} \frac{m}{1 + e^{-2mx_1}},
\label{domain_wall}
\end{eqnarray}
which also satisfies the Gauss' law (\ref{gauss1}) and (\ref{gauss2}).
This solution is just the one first found in \cite{HaLi}, 
but they obtained it in the D-term deformed theory.

We can see that the solution 
 (\ref{domain_wall}) keeps a half of $\mathcal{N} = 2$ supersymmetry 
 specified by
\begin{eqnarray}
\mathcal{Q}_{\pm} \equiv Q_1 \pm i \bar{Q}_2,
\end{eqnarray}
while the one in \cite{HaLi} keeps half of the maximal $\mathcal{N} = 6$ supersymmetry.
The tension of the domain wall is evaluated as 
\begin{eqnarray}
T = N (N-1) \frac{km^2}{4\pi} \sim m^2 N^2, \quad (N \to \infty).
\label{tension_large_N}
\end{eqnarray}
This agrees with the result (\ref{dw_tension}).

Before going to the next section, we would like to consider 
the reducible vacuum solutions discussed in \cite{GoRoRaVe}. 
Because the BPS matrices $S^A$ for any partition of $N$ do satisfy 
the vacuum condition, there is a set of reducible 
solutions that is obtained by replacing $S^A$ by 
\begin{eqnarray}
\tilde{S}^A = 
\left(
\begin{array}{ccc}
S^A_{N_1} & & \\
 & \cdots & \\
 & & S^A_{N_l}
\end{array}
\right).
\end{eqnarray}
Here $N_b, \ (b=1, \cdots, l)$ satisfy $\sum^{l}_{b=1} N_b = N$ 
and $S^A_{N_b}$ are the BPS matrices in an $N_b$ dimensional representation.
We will see 
in a later section 
 that these vacua correspond to a set of fuzzy $S^3$ shells 
 with different values of the radius 
 and that the potential is written as in (\ref{pot-M2-shell}). 

\begin{figure}[t]
\begin{center}
\includegraphics[scale=.7]{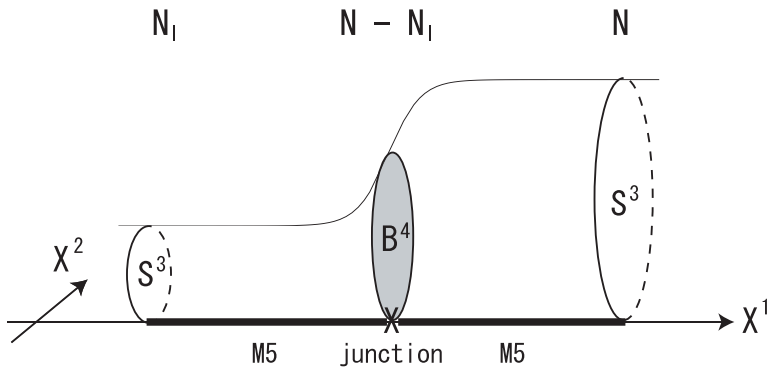}
\end{center}
\caption{A schematic picture of a domain wall solution.}
\label{domain_wall_pic}
\end{figure}

It is possible to capture the brane configuration corresponding to a 
domain wall interpolating between one of the reducible vacua constructed,
for example, 
 by $\tilde{S}^A = \mathrm{diag} (S^A_{N_l}, 0_{N-N_l}), \ N_l < N$ at $x_1 = - \infty$ 
and the irreducible vacuum with dimension $N$ at $x_1 = + \infty$.
It is easy to show that the large-$N$ behavior of its tension is the same as in (\ref{tension_large_N}).
Following the discussion in \cite{PoSt,Be}, in the region $x_1 < 0$, 
the configuration is interpreted as an M5-brane extending in the $(x^0,x^1,x^2)$ 
directions and wrapping an $S^3$ with 
M2-brane charge $N_l$ while in the region $x_1 > 0$, it is an M5-brane 
extending along the same directions and wrapping an $S^3$ with charge $N$. 
They are generically both bent and meet at a point $x_1 = 0$. Because of charge conservation, 
another new 
M5-brane with charge $N-N_l$ should exist at that point. This is a kind of brane junction. 
Because all M5-branes meet at a point, one direction should be 
supplemented for the new M5-brane.
This is possible if we consider a ball $B^4$ instead of a 
sphere $S^3$.
Therefore the other new M5-brane is extending along the 
$(x^0,x^2)$ directions and filling a four-ball $B^4$ with its surface $S^3$.
A schematic picture is given in Fig. \ref{domain_wall_pic}.

\section{Effective Hamiltonian and BPS equations}
In this section, motivated by the discussion in the previous section on 
the spherical M5-brane at the vacuum, 
we study the effective Hamiltonian 
for the configuration of polarized M2-branes.
We will see that the potential 
for the radius of this configuration coincides with 
(\ref{radial_potential}) at large $N$.
Consider an ansatz
\begin{eqnarray}
& & Z^A = W^{\dagger A} = f(x) S^A, \quad Z^{\dagger}_A = W_A = \bar{f} 
(x) S^{\dagger}_A, \ f(x) \in \mathbf{C}, \nonumber \\
& & A_{\mu} = a_{\mu} (x) S^B S^{\dagger}_B, \quad \hat{A}_{\mu} = 
a_{\mu} (x) S^{\dagger}_B S^B, \ a_{\mu} (x) \in \mathbf{R},
\label{ansatz}
\end{eqnarray}
where $S^A$ are the BPS matrices. 
This configuration represents the M2-branes polarized into a fuzzy $S^3$.
The physical radius of the fuzzy $S^3$ is given by 
\begin{eqnarray}
R^2 = \frac{2(N-1)}{T_2} |f (x)|^2.
\end{eqnarray}
Under the ansatz, the covariant derivative becomes
\begin{eqnarray}
D_{\mu} Z^A 
= (\partial_{\mu} f (x) + i a_{\mu} f (x)) S^A \equiv (\mathcal{D}_{\mu} f (x)) S^A,
\end{eqnarray}
while the gauge field strength satisfies
\begin{eqnarray}
F_{\mu \nu} &=& (\partial_{\mu} a_{\nu} - \partial_{\nu} a_{\mu}) S^B 
S^{\dagger}_B \equiv f_{\mu \nu} S^B S^{\dagger}_B, \\
\hat{F}_{\mu \nu} &=& f_{\mu \nu} S^{\dagger}_B S^B. 
\end{eqnarray}
The equation of motions for $A_{\mu}$ and $\hat{A}_{\mu}$ reduce to 
 the following equation for $a_{\mu}$,
\begin{eqnarray}
\frac{k}{4\pi} \epsilon^{\rho \mu \nu} f_{\mu \nu} 
= 2 i (f \mathcal{D}^{\rho} \bar{f} - \bar{f} \mathcal{D}^{\rho} f).
\end{eqnarray}
For the case of the ansatz (\ref{ansatz}), we have
\begin{eqnarray}
V_D = 0, \quad 
\frac{\partial \mathcal{W}}{\partial Z^A} 
= - \left( \frac{k m}{8 \pi} - \frac{1}{2} |f|^2 \right) \bar{f} 
S^{\dagger}_A, \quad 
\frac{\partial \mathcal{W}}{\partial W_A} 
= - \left( \frac{k m}{8 \pi} - \frac{1}{2} |f|^2 \right) f S^A.
\end{eqnarray}
Therefore, from the equation (\ref{ABJM_hamiltonian}), 
 the effective Hamiltonian for 
the polarized M2-branes is given by
\begin{eqnarray}
H &=& 
2 N (N-1) \int \! d^3 x \ 
\left[
\mathcal{D}_0 f \mathcal{D}_0 \bar{f} + \mathcal{D}_i f \mathcal{D}_i 
\bar{f} + \frac{64 \pi^2}{k^2} |f|^2 \left(\frac{km}{8\pi} - \frac{1}{2} 
|f|^2 \right)^2 
\right].
\label{effective_hamiltonian}
\end{eqnarray}
This is nothing but the Hamiltonian for 
an abelian Chern-Simons Higgs model studied in 
\cite{JaLeWe, HoKiPa, JaWe} and \cite{KiLeYe}.
Note that all the non-abelian structures in the model have been encoded into 
the overall factor $N (N-1)$ by the help of the BPS matrices.
Although the gauge symmetry has been effectively reduced to the abelian 
symmetry, the Hamiltonian still describes $N$ M2-branes.

It is known that the abelian Chern-Simons-Higgs model with the potential 
given in (\ref{effective_hamiltonian}) 
exhibits a three-dimensional $\mathcal{N} = 2$ supersymmetry \cite{LeLeWe}
and admits various class of BPS soliton solutions.
See appendix C for the full structure of the Lagrangian 
including fermion parts.
Following the analysis in \cite{KiLeYe}, 
we find a Bogomol'nyi completion of the energy,
\begin{eqnarray}
H &=& 2 N (N-1) \int \! d^3 x \ 
\left[
\left| 
\mathcal{D}_0 f \pm i f \left( m - \frac{4 \pi}{k} |f|^2 \right) \cos 
\alpha \pm \mathcal{D}_2 f \sin \alpha
\right|^2 \right. \nonumber \\
& & \left. \qquad \qquad \qquad \qquad
+ \left|
\mathcal{D}_1 f \mp i \mathcal{D}_2 f \cos \alpha 
\mp f 
\left( 
m - \frac{4 \pi}{k} |f|^2 
\right)
\sin \alpha
\right|^2 
\right] \nonumber \\
& & \qquad \qquad \qquad \qquad \pm \mathbf{B} \cos \alpha \mp \mathbf{S} \cos \alpha 
\mp \mathbf{P} \sin \alpha \pm \mathbf{T} \sin \alpha.
\label{Hamiltonian}
\end{eqnarray}
where we have used the gauge field equation of motion.
Therefore we have an energy 
bound
\begin{eqnarray}
E \ge
 2 N (N-1) \sqrt{(\mathbf{B} - \mathbf{S})^2 + (\mathbf{P} - \mathbf{T})^2}.
\label{energy_bound}
\end{eqnarray}
Here the following 
quantities have been defined,
\begin{eqnarray}
& & \mathbf{B} \equiv \frac{km}{4\pi} \int \! d^3 x \ B, \quad B\equiv
 f_{12}, \\
& & \mathbf{S} \equiv i \int \! d^3 x \ 
\left[
\partial_1 (f \mathcal{D}_2 \bar{f}) - \partial_2 (f \mathcal{D}_1 \bar{f})
\right], \\
& & \mathbf{P} \equiv \int \! d^3 x \ 
\left[
\mathcal{D}_0 f \mathcal{D}_2 \bar{f} + \mathcal{D}_2 f \mathcal{D}_0 \bar{f}
\right], \\
& & \mathbf{T} \equiv \int \! d^3 x \ 
\mathcal{D}_1 
\left( 
m |f|^2 - \frac{2\pi}{k} |f|^4
\right).
\end{eqnarray}
Here $\mathbf{B}, \mathbf{S}, \mathbf{P}, \mathbf{T}$ are 
(time integrals of) the magnetic flux, the angular momentum, the linear momentum 
along $x^2$ direction, the tension of a domain wall. 
The energy bound (\ref{energy_bound}) is saturated when the 
angle $\alpha$ satisfies the following condition
\begin{eqnarray}
\cos \alpha = \frac{\mathbf{B} - \mathbf{S}}{\sqrt{(\mathbf{B} - 
\mathbf{S})^2 + (\mathbf{T} - \mathbf{P})^2}}, \quad 
\sin \alpha = \frac{\mathbf{T} - \mathbf{P}}{\sqrt{(\mathbf{B} - 
\mathbf{S})^2 + (\mathbf{T} - \mathbf{P})^2}}.
\end{eqnarray}

The baryonic $U(1)$ charge (\ref{baryonic_charge}) 
for our ansatz is evaluated as 
\begin{eqnarray}
Q_b = 2 i N(N-1) \int \! d^2 x \ 
\left[
f \mathcal{D}_0 \bar{f} - \bar{f} \mathcal{D}_0 f
\right].
\end{eqnarray}
Using the equation of motion for the gauge field, we have 
the relation
\begin{eqnarray}
Q_b = 
- \frac{k}{2\pi} N (N-1) \Phi, \quad 
\Phi \equiv \int \! d^2 x \ B.
\label{charge_flux}
\end{eqnarray}
The Noether charge is proportional to the  magnetic flux.
This is a specific property of Chern-Simons solitons.
From the Bogomol'nyi completed form of the Hamiltonian 
(\ref{Hamiltonian}), we obtain the BPS equations,
\begin{eqnarray}
& & \mathcal{D}_0 f \pm i f \left( m - \frac{4 \pi}{k} |f|^2 \right) \cos 
\alpha \pm \mathcal{D}_2 f \sin \alpha = 0, \label{BPS01} \\ 
& & \mathcal{D}_1 f \mp i \mathcal{D}_2 f \cos \alpha 
\mp f 
\left( 
m - \frac{4 \pi}{k} |f|^2 
\right)
\sin \alpha = 0. \label{BPS02}
\end{eqnarray}
Requiring that the supersymmetry transformations (\ref{susy-trans1}) and (\ref{susy-trans2})
of the fermionic fields vanish combined with these BPS equations yields 
the following conditions,
\begin{eqnarray}
&& \epsilon_\alpha \pm
  \gamma^0_{\alpha\beta}\bar{\epsilon}^\beta\cos\alpha
  \pm i\gamma^1_{\alpha\beta}\bar{\epsilon}^\beta \sin\alpha=0,\\
&& \mp i\gamma^0_{\alpha\beta}\bar{\epsilon}^\beta\sin\alpha
  \mp\gamma^1_{\alpha\beta}\bar{\epsilon}^\beta\cos\alpha
  +i\gamma^2_{\alpha\beta}\bar{\epsilon}^\beta=0,
\end{eqnarray}
where $\gamma^\mu_{\alpha\beta}$ are the three-dimensional $\gamma$-matrices given
 in appendix B.
This tells us that the BPS equations retain a quarter of the $\mathcal{N} = 2$
 supersymmetry, specified by the effective supercharge 
 (in the case $\cos\alpha=1$) 
\begin{eqnarray}
 {\cal Q}=Q_1 \pm iQ_2 +i\bar{Q}_1 \mp \bar{Q}_2\,.
\end{eqnarray} 

One should be careful about the above procedure deriving the BPS 
equations (\ref{BPS01}), (\ref{BPS02}).
We first assumed an ansatz and derived the effective Hamiltonian 
(\ref{Hamiltonian}) which is valid only for the special configuration (\ref{ansatz}). 
There is a possibility that 
the BPS equations are not consistent with the original equations of motion.
Let us check this issue.
Once the equations (\ref{BPS01}), (\ref{BPS02}) are satisfied, we have
\begin{eqnarray}
\mathcal{D}_{\mu} \mathcal{D}^{\mu} f = 
m^2 f 
- \frac{16\pi m}{k} |f|^2 f + 3 \left(\frac{16\pi^2}{k^2}\right) |f|^4 f.
\end{eqnarray}
This is the equation of motion (\ref{eom}) 
for $Z^A = W^{\dagger A}$ for the ansatz (\ref{ansatz}). 
Therefore the BPS equations (\ref{BPS01}), (\ref{BPS02}) are consistent 
with the full (second order) equations of motion for $Z^A, W_A$.

\begin{figure}[t]
\begin{center}
\includegraphics[scale=.4]{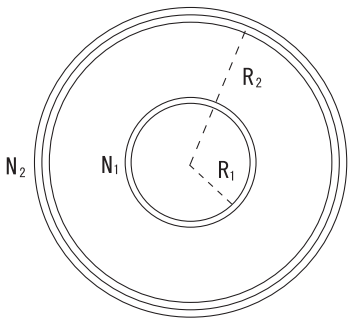}
\end{center}
\caption{A schematic picture of the reducible fuzzy $S^3$ shell configuration. 
$l = 2$ case.
A number of $N_1 = 2$, $N_2 = 3$ M2-branes are polarized around $R_1 \sim \sqrt{N_1}$ and $R_2 
\sim \sqrt{N_2}$. 
}
\label{shell}
\end{figure}

Note that in terms of the physical fuzzy $S^3$ radius $R$, 
 the potential term in the equation (\ref{Hamiltonian}) becomes
\begin{eqnarray}
V(R) = 
{4\pi^2 N T_2^3 R^2 \over k^2(N-1)^2}\left(R^2 - {(N-1)km \over 2\pi T_2}\right)^2.
\end{eqnarray}
At large $N$, this gives 
\begin{eqnarray}
V(R) \sim T_5 \frac{T_2}{N}
\left(
R^3 - \frac{k}{4\pi}
\frac{m N}{T_2} R
\right)^2,
\end{eqnarray}
where we have used $T_p=(2\pi)^p M_{11}^{-p+1}$.
After a trivial rescaling of $R$, the result agrees 
with the potential (\ref{radial_potential}) 
evaluated from the spherical M5-brane.
Once we consider the reducible ansatz $S^A \to \tilde{S}^A$, 
the potential is a sum of the each $S^A_{N_b}$ sectors, {\it i.e.}, it is 
given by
\begin{eqnarray}
V(R) \sim \sum_{b = 1}^l T_5 \frac{T_2}{N_b}
\left(
R_b^3 - \frac{k}{4\pi}
\frac{m N_b}{T_2} R_b
\right)^2, \quad \sum_{b=1}^l N_b = N.
\label{reducible_potential}
\end{eqnarray}
This has the same form of the potential for the 
 M2-branes distributed over a 3-sphere 
(\ref{pot-M2-shell})
 giving a 3-sphere shells of M2-branes \cite{Be}. 
Therefore the reducible configuration can be interpreted as a set of 
$N_b$ M2-branes polarized into fuzzy $S^3$ spheres with radii $R_b$.
See Fig. \ref{shell} for a schematic picture.

A comment on this reducible configuration is in order.
We have ignored off-block-diagonal parts of $\tilde{S}^A$ to derive the 
equation (\ref{reducible_potential}). The result is the sum of the potentials 
for the radii $R_b$ inside $N_b$ stack of M2-branes 
without any interactions among each stacks.
Once we turn on the off-block-diagonal parts, it represents interactions 
among each stack of M2-branes and hence, in the dual picture, 
interactions among M5-branes.

\section{Solutions}
In this section, we explore solutions for the BPS equations (\ref{BPS01}), 
 (\ref{BPS02}). 
Since it is difficult to find solutions for general cases, we focus on 
the simplest situations.
Let us first consider the $\cos \alpha = 1$ static case. In this case 
$(\mathbf{T} - \mathbf{P})$ must vanish and 
the energy bound is given by $\mathbf{B}$ and $\mathbf{S}$.
The BPS equations reduce to 
\begin{eqnarray}
& & \mathcal{D}_1 f \mp i \mathcal{D}_2 f = 0, \label{static_BPS2} \\
& & - \frac{k}{8\pi} B 
 \pm |f|^2 \left(m - \frac{4\pi}{k} |f|^2 \right) 
= 0. \label{static_BPS1}
\end{eqnarray}
This is nothing but the equation (14) in \cite{HoKiPa}. 
Following \cite{JaLeWe}, we now assume an ansatz 
\begin{eqnarray}
f (r, \theta) &=& \lambda g (r) e^{i n \theta}, \ \lambda^2 = 
\frac{km}{4\pi}, \
g (r) \in \mathbf{R}, \label{ansatz1} \\
a_i (r) &=& \epsilon_{ij} \frac{\hat{x}^j}{r} [a (r) - n] \ (i=1,2), \label{ansatz2}
\end{eqnarray}
where $(r,\theta)$ are polar coordinates in the M2-brane world-volume and $n$ 
is an integer. The BPS equations (\ref{static_BPS2}), (\ref{static_BPS1}) reduce to 
\begin{eqnarray}
g'(r) &=& \mp \frac{1}{r} a (r) g (r), \label{rBPS1} \\
\frac{a'(r)}{r} &=& \mp 2 m^2 g^2 (r)(g^2 (r) - 1). \label{rBPS2}
\end{eqnarray}
Here the prime stands for the differentiation with respect to $r$. 
Requiring $a_i (r)$ and $g (r)$ to be non-singular at the origin, we obtain 
boundary conditions
\begin{eqnarray}
a (0) - n = 0, \quad n g (0) = 0. \label{bd1}
\end{eqnarray}
On the other hand, the condition of finite energy is 
 requiring that the scalar field settles down to its vacuum 
 configuration at infinity, namely, 
\begin{eqnarray}
g (\infty) = 
\left\{
\begin{array}{l}
0 \quad \textrm{(symmetric phase)}, \\
1 \quad \textrm{(broken phase)}.
\end{array}
\right. \label{bd2}
\end{eqnarray}
From the equations (\ref{rBPS1}), (\ref{rBPS2}), we obtain
\begin{eqnarray}
(\ln g^2)'' + \frac{1}{r} (\ln g^2)' + 4 m^2 g^2 (1 - g^2) = 0.
\end{eqnarray}
When $g$ is small, the $\mathcal{O} (g^4)$ term can be neglected and 
 we can find an explicit solution \cite{JaLeWe}. 
That is given by
\begin{eqnarray}
g (r) = \frac{\sqrt{8} s}{2 m r} 
\left[
\left(\frac{r}{r_0} \right)^s
+ \left(\frac{r_0}{r} \right)^s
\right]^{-1}.
\end{eqnarray}
Here $r_0, s$ are arbitrary constants. 
The energy density for the ansatz (\ref{ansatz1}) and
 (\ref{ansatz2}) is given by
\begin{eqnarray}
E&=&\int_0^\infty dr {\cal E},\\
{\cal E}&=& 4\pi r N(N-1)\lambda^2\left(
{(g')^2+{g^2a^2 \over r^2}+\left(a^\prime \over 2mrg\right)^2+m^2g^2(g^2-1)^2}
\right).
\end{eqnarray}
The magnetic charges and angular momentum are also readily written down
 as
\begin{eqnarray}
 && \Phi ={km \over 2}\int dr a'(r), \label{mag} \\
 && S =2\pi \lambda^2 \int dr (2a(x)g(x)g'(x)+a'(x)g^2(x)), \label{ang}
\end{eqnarray}
and ${\bf P}=0$ and ${\bf T}=0$. 
Here we have defined $\mathbf{S} \equiv \int \! dt \ S$.
Considering the boundary conditions (\ref{bd1}) and (\ref{bd2}),
(\ref{mag}) and (\ref{ang}) become
\begin{eqnarray}
 \Phi = -{km \over 2}(n+\alpha), \quad S=-2\pi\lambda^2 \alpha\beta\,
\end{eqnarray}
where $\alpha=-a(\infty)$ and $\beta=g(\infty)$.
In the following analysis, we consider the case $B<0$, which corresponds
 to choosing the plus sign in (\ref{ansatz1}) and (\ref{ansatz2}).
We will also see that $S=0$ for all the cases we will discuss below.
Because 
the detailed study of the global structure of these solutions has been
 performed in \cite{JaLeWe},
we just briefly describe the solutions according to the boundary conditions.

\subsection{Vortices}
\begin{figure}[t]
\begin{center}
$
\begin{array}{ccc}
  \epsfxsize=7cm
  \epsfbox{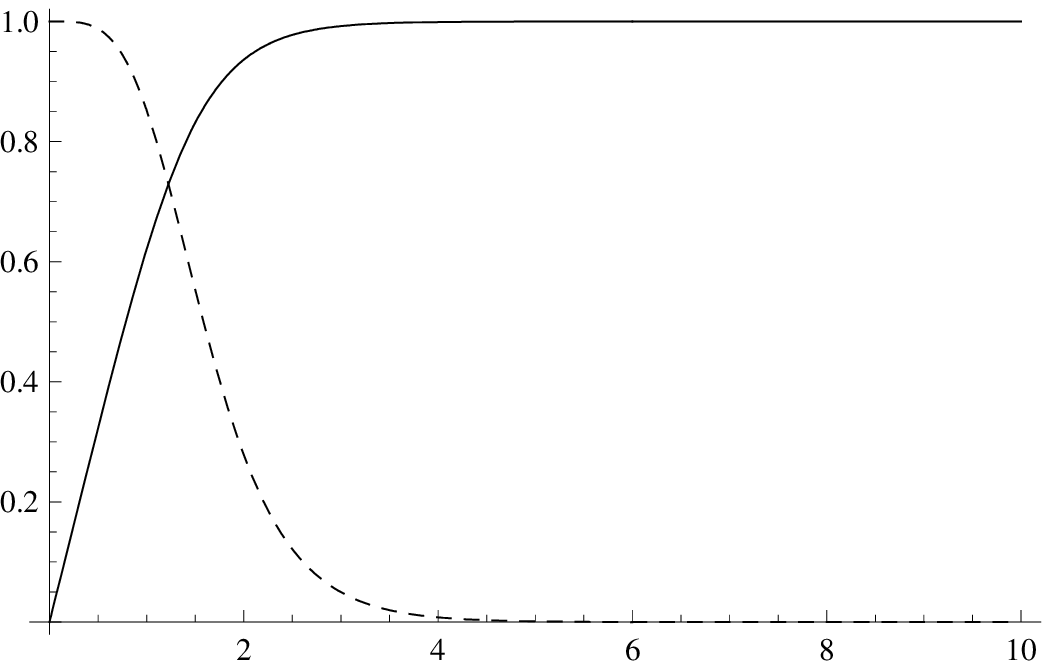} 
 & &
  \epsfxsize=7cm
  \epsfbox{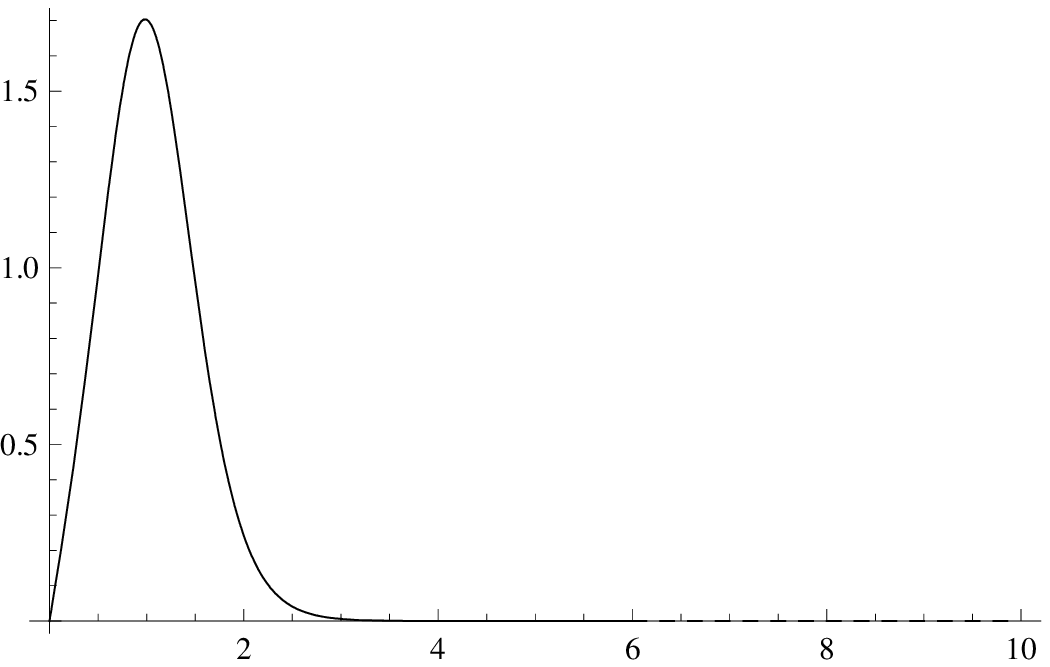} \\
 \end{array}
$
\caption{The left figure shows the profiles of the function $g(r)$ (solid line) and
 $a(r)$ (dashed line) for the topological vortex with $m=1$.
 The right figure shows the behavior of
 corresponding energy density with $k=1$ and $N=2$.}
 \label{vortex}
\end{center}
\end{figure}

\begin{figure}[t]
\begin{center}
  \epsfxsize=8cm
  \epsfbox{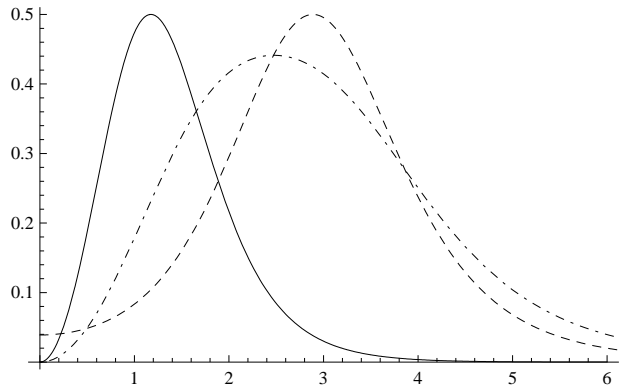} 
\caption{Plots of 
the absolute value of
 the magnetic flux density with $k=1$ and $m=1$. Solid,
 dashed, dot-dashed lines correspond to vortex, 
Q-ball and non-topological vortex, respectively.}
 \label{mf}
\end{center}
\end{figure}

First consider the case of the boundary condition $g (\infty) = 1$. 
For the $n=0$ case, the only solution is the vacuum configuration, $g(r)=1$ and $a(r)=0$.
Thus it is just the spherical M5-brane discussed in section 3.
For the $n\not=0$ case, a topologically non-trivial configuration appears,
 which connects the boundaries $g(0)=0, a(0)=n$ and $g(\infty)=1, a(\infty)=0$. 
They are vortex solutions. 
Although, no explicit solutions of the equation for this 
boundary condition is not known, it is possible to solve it numerically.
In Fig. \ref{vortex}, we plot $g(r)$ and $a(r)$ for the $n=1$ solution
 and its corresponding energy density.
We also show the magnetic flux density in Fig. \ref{mf}.
We can see that the topological charge is given by the 
 magnetic flux 
$ \Phi = - \frac{kmn}{2}$. 
As can be seen in Fig. \ref{vortex}, 
in almost all of the region of the M2-brane world-volume, the configuration represents the 
vacuum fuzzy $S^3$ 
discussed in section 3 
but at the origin, there is a ``dimple'' inside which the 
full supersymmetry is recovered.

%%%%%%%%%%%%%%%%%%%%%%%%%%%%%%%%%%%%%%%%%%%%%%
%  Vortex
%%%%%%%%%%%%%%%%%%%%%%%%%%%%%%%%%%%%%%%%%%%%%%
%%%%%%%%%%%%%%%%%%%%%%%%%%%%%%%%%%%%%%%%%%%%

\subsection{Q-balls}
\begin{figure}[t]
\begin{center}
$
\begin{array}{ccc}
  \epsfxsize=7cm
  \epsfbox{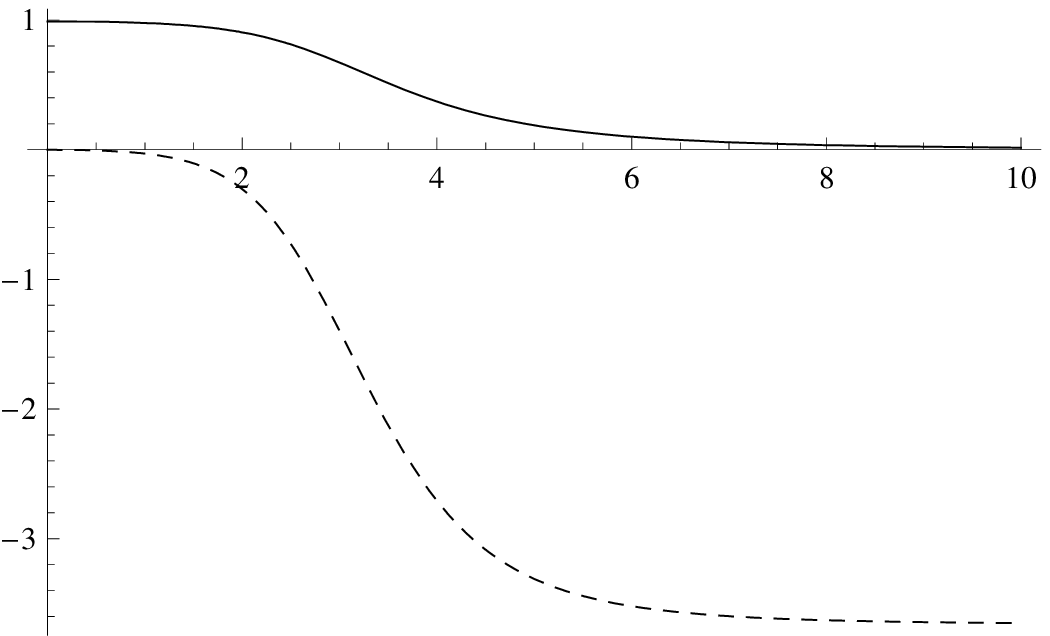} 
 & &
  \epsfxsize=7cm
  \epsfbox{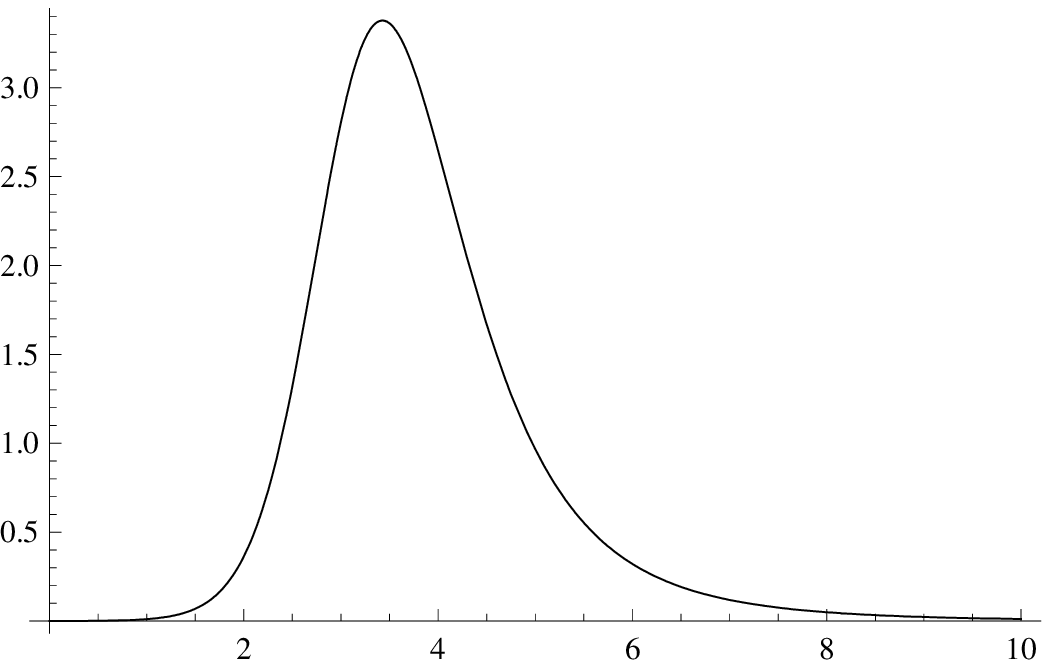} \\
 \end{array}
$
\caption{The left figure shows the profiles of the function $g(r)$ (solid line) and
 $a(r)$ (dashed line) 
for the Q-ball solution 
with $m=1$. The right figure shows the behavior of
 corresponding energy density with $k=1$ and $N=2$.}
 \label{lump}
\end{center}
\end{figure}

Next we consider the case $g(\infty)=0$ and $n=0$.
In this case, the scalar function $f(x)$ approaches the trivial
 vacuum and therefore all configurations are topologically trivial.
The gauge field part $a(r)$ starts from the boundary 
$a(0)=0$ and approaches some value $a(\infty)$ which characterizes the value of
 magnetic flux $\Phi$.  
The magnetic flux is given by the $U(1)$ Noether charge 
through the relation (\ref{charge_flux}).
This kind of solution is called a Q-ball .
This is a lump-like object localized on a ring which surrounds 
the origin. 
Around the origin, the fuzzy $S^3$ has non-zero radius but the full 
supersymmetry is generically broken there.
This is because the value of $g (r)$ at the origin need not to be the 
value in the vacuum. The situation is similar at infinity where the 
fuzzy sphere collapses into zero size but the gauge field has non-zero value.
The plots of the solution 
and corresponding energy density are shown in Fig. \ref{lump}.
The behavior of the magnetic flux is shown in Fig. \ref{mf}.

%%%%%%%%%%%%%%%%%%%%%%%%%%%%%%%%%%%%%%%%%%%%%%
%  Lump
%%%%%%%%%%%%%%%%%%%%%%%%%%%%%%%%%%%%%%%%%%%%%%
%%%%%%%%%%%%%%%%%%%%%%%%%%%%%%%%%%%%%%%%%%%%

\subsection{Non-topological vortices}
\begin{figure}[t]
\begin{center}
$
\begin{array}{ccc}
  \epsfxsize=7cm
  \epsfbox{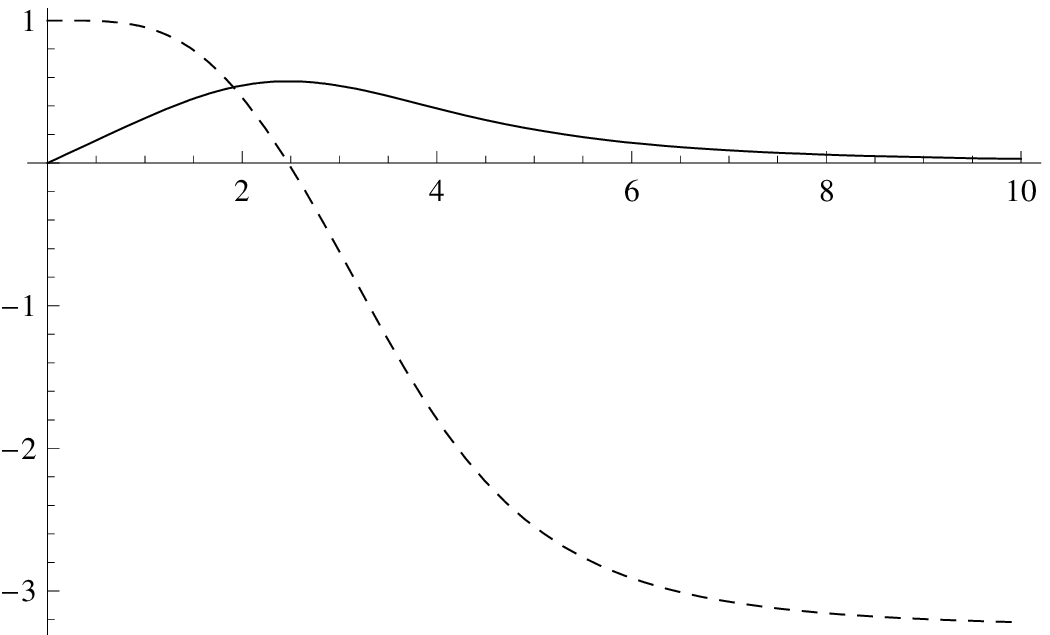} 
 & &
  \epsfxsize=7cm
  \epsfbox{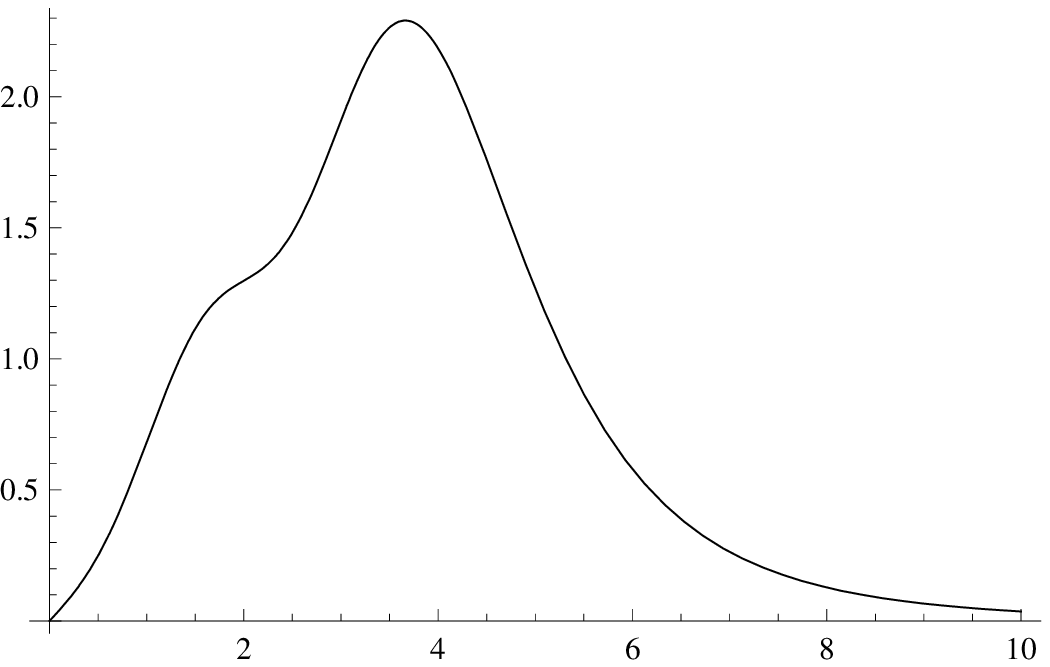} \\
 \end{array}
$
\caption{The left figure shows that the profiles of the function $g(r)$ 
(solid line) and  $a(r)$ (dashed line) with $m=1$. 
The right figure shows the behavior of corresponding energy density 
for the non-topological vortex with $k=1$ and $N=2$.}
\label{non-vortex}
\end{center}
\end{figure}

Here we consider the case $g(\infty)=0$ and $n\neq 0$.
The solutions in this case are hybrids of the previous two cases.
The large distance behaviors of $g(r)$ and $a(r)$ are the same as
 those in the Q-ball case, while they are similar to the vortex solution 
around the origin.
These types of solutions are called non-topological vortices. 
Again, around the origin, the supersymmetry is recovered.
We show the plots of this type of solution for the $n=1$ case,
 its energy density in Fig. \ref{non-vortex} and the flux density in
 Fig. \ref{mf}.
%%%%%%%%%%%%%%%%%%%%%%%%%%%%%%%%%%%%%%%%%%%%%%
%  Non-topological vortex
%%%%%%%%%%%%%%%%%%%%%%%%%%%%%%%%%%%%%%%%%%%%%%
%%%%%%%%%%%%%%%%%%%%%%%%%%%%%%%%%%%%%%%%%%%%

\subsection{Gauged domain wall}
Next, let us consider $\cos \alpha = 1$ but non-static case. The BPS equations 
(\ref{BPS01}), (\ref{BPS02}) reduce to 
\begin{eqnarray}
& & \mathcal{D}_0 f \pm i f \left(m - \frac{4 \pi}{k} |f|^2 \right) = 0, 
\label{BPS11} \\
& & \mathcal{D}_1 f \mp i \mathcal{D}_2 f = 0. \label{BPS12}
\end{eqnarray}
Considering the upper sign in the equations and assuming the following ansatz 
\begin{eqnarray}
& & f (t, x_1, x_2) = \phi (x_1) e^{- i m (t + x_2)}, \quad \phi (x_1) \in 
\mathbf{R}, \\
& & a_0 (x_1) = a_2 (x_1), \quad a_1 = 0,
\end{eqnarray}
the equation (\ref{BPS11}) implies
\begin{eqnarray}
a_0 = a_2 = \frac{4\pi}{k} \phi^2 (x_1).
\end{eqnarray}
On the other hand, the equation (\ref{BPS12}) reduces to
\begin{eqnarray}
\partial_1 \phi - m \phi + \frac{4\pi}{k} \phi^3 = 0 \label{eq_phi}.
\end{eqnarray}
The gauge field equations of motion are satisfied provided $\phi$ 
satisfies the equation (\ref{eq_phi}). 
Solutions of (\ref{eq_phi}) are 
\begin{eqnarray}
\phi^2 (x_1) &=& \frac{k}{4\pi} 
\frac{m}{1 \pm A e^{-2 m x_1}},
\end{eqnarray}
where $A>0$ is an integration constant. 
This solution was first found in \cite{KiLeYe} and was there called a ``supertube'' 
solution. 
The solution which has plus sign 
in front of $A = 1$ corresponds to a domain wall which interpolates between 
$\phi^2 (-\infty) = 0$ and $\phi^2 (+ \infty) = \frac{km}{4\pi}$. 
At $x_1 = + \infty$, it describes an M5-brane wrapping an $S^3$ while 
$x_1 = - \infty$, is the trivial vacuum. 
Although, we are considering the $\cos \alpha = 1$ case, and hence 
$\mathbf{T} - \mathbf{P} = 0$, 
each $\mathbf{T}$ and $\mathbf{P}$ have non-vanishing 
values. 
Indeed, the tension and the momentum for the solution 
are the same value $T = P = \frac{km^2}{8\pi}$ and 
$\mathbf{S} = 0$, where $\mathbf{T} = \int \! d x^0 d x^2 \ T$
 and $\mathbf{P} = \int \! d x^0 d x^2 \ P$.
This fact means that the domain wall 
tension is canceled by the momentum
and makes it bend freely \cite{KiLeYe}.

This solution would be interpreted as a brane configuration 
by following the discussion in section 3.
In the solution, $\phi (x_1)$ part corresponds to the 1/2 BPS 
domain wall solution (\ref{domain_wall}). 
An M5-brane at $x_1 > 0$ bends and an M5-brane filling $B^4$ 
may appear 
at the origin. On the other hand, the phase factor $e^{-im (t + x_2)}$ 
acts as a rotation in the 3-7, 4-8, 5-9, 6-10 plane \cite{Be}. 
This can be seen by decomposing $Z^A = X^{A+2} + i X^{A+6}, \ W^{\dagger 
A} = X^{A+4} + i X^{A+8} $ where the $X$ and $i X$ parts correspond to 
hermitian and anti-hermitian parts of the bi-fundamental scalar fields.
Therefore, the solution can be interpreted as a configuration 
of the bent branes rotating in time and its angle also depends on $x_2$. 

\section{Conclusions and discussions}
In this paper, we investigated the ABJM model deformed by an F-term mass. 
The mass term generically breaks $\mathcal{N} = 6$ maximal supersymmetry 
down to $\mathcal{N} = 2$. 
A vacuum of the model corresponding to an irreducible 
representation is found to be a fuzzy $S^3$ representing an M5-brane 
with topology $\mathbf{R}^{1,2} \times S^3$ at large-$N$. 
We find that the radius of the sphere coincides with the one found in 
\cite{Be}. The tension of the domain wall solutions is evaluated and 
compared with the result of the dual M5-brane calculation.
With this observation, we identified the F-term mass $m$ 
as the VEV of the anti-self-dual 4-form flux $T_4$ in the 
eleven-dimensional supergravity that was introduced in \cite{Be}. 
Therefore the vacuum configuration can be interpreted as Myers' 
dielectric effect making $N$ M2-branes puff up into a fuzzy sphere.

In the latter half of this paper, 
we assumed a polarized M2-brane configuration.
We showed that the mass deformed ABJM model reduces 
to the abelian Chern-Simons-Higgs model with 6th power polynomial 
potential studied 
in detail 
by Jackiw-Lee-Weinberg \cite{JaLeWe}.
The potential part of this effective model is just the potential for the 
radius of the spherical M5-brane. 
This effective model admits not only BPS topological vortices and 
domain walls but also non-topological vortices and Q-balls.
We find that these BPS topological/non-topological solitons preserve 1/4 of 
the manifest $\mathcal{N} = 2$ supersymmetry. Because these solutions can exist only 
in the $m \not= 0$ case, these configurations are supported against 
collapse by the non-zero background flux $T_4$. 
The situation is quite similar to the massive BLG model where various 
BPS soliton solutions are possible \cite{HoLeLe}.

A comment on the number of supersymmetries is in order. 
In the massless ABJM model, $SU(2) \times SU(2)$ global symmetry 
together with $SU(2)_R$ symmetry gives $SU(4)_R \sim SO(6)_R$ 
implying the existence of $\mathcal{N} = 6$ supersymmetry. 
Indeed, in \cite{Te}, the explicit on-shell $\mathcal{N} = 6$ 
supersymmetry transformation of component fields was found.
Our case is similar to the massless ABJM model.
The global $SU(2) \times SU(2)$ symmetry is broken down to 
$SU(2)_{\mathrm{diag}}$ by the mass term. However, this 
$SU(2)_{\mathrm{diag}}$ does not commute with the remaining $SU(2)_R$. These 
two $SU(2)$s are combined into an $SO(4)_R$ which would imply an 
enhancement of $\mathcal{N} = 2$ supersymmetry to $\mathcal{N} = 4$. 
Presumably, this supersymmetry is enhanced to the maximal one since 
our deformation term (\ref{mass_superpot}) (together with its conjugate term) does not break the $U(1)_R$ and 
$\mathbf{Z}_2$ symmetries and these are combined with the $SU(2)$s giving $SU(2) 
\times SU(2) \times U(1)_R \times \mathbf{Z}_2$ symmetry as discussed in \cite{GoRoRaVe}.
Although we do not present it here, it would be possible to 
explicitly find the enhanced supersymmetries by the Noether 
method or requiring that the supersymmetry algebra closes.

Another important issue is the understanding of the solution in the dual 
spherical M5-brane world-volume picture. Because our effective potential 
is identified with the radius potential of the spherical M5-brane in the 
presence of the anti-self-dual 4-form flux, the BPS solutions 
discussed in this paper correspond to BPS configurations of this 
 spherical M5-brane. It seems possible to find these BPS configurations 
 by promoting the constant radius $r$ in (\ref{radial_potential}) 
to a ``field'' $r = 
 r(x_1,x_2)$ on the M2-brane world-volume and solving the BPS equations 
 obtained from the promoted spherical M5-brane action. 
This issue is beyond the scope of this paper and we leave it for future 
work.

Finally, let us comment on the dimensional reduction of the ABJM model.
Since the ABJM model describes $N$ coincident M2-branes in eleven dimensions, 
once we compactify one direction along the M2-brane world-volume, 
the model should describe the low energy effective theory of $N$ 
coincident fundamental strings (F1) in type IIA string theory.
This theory is similar to the matrix string theory \cite{DiVeVe}  
representing a polarization of the F1 due to the background flux.
The configuration corresponding to the M2-branes polarized into the 
M5-brane now reduces to fundamental strings polarized into a D4-brane
\footnote{A similar configuration for F1 polarized into D2 was studied in the dilute set of 
D0-branes having the F1 charge \cite{Be2}.}.
The anti-self-dual 4-form flux in eleven-dimensions is reduced to the 
anti-self-dual RR 4-form flux in ten-dimensions. 
The multiple F1 is polarized into a fuzzy $S^3$ due to this flux.
Since the potential terms both in the M5-brane action and the massive 
ABJM model do not change under the dimensional reduction, the discussion 
in section 3 holds even for the D4/F1 case implying $AdS_3/CFT_2$ duality.

On the other hand, the effective Lagrangian for the polarized M2-branes 
discussed in section 4 reduces to the one for the polarized F1. 
One component of the gauge field $a_{\mu}$ becomes an auxiliary field after the 
dimensional reduction. The resultant theory is a new two dimensional BF-like theory.
We find that this model is exactly the one studied in \cite{KaLeLe}. 
It was shown that the model exhibits $\mathcal{N} = (2,2)$ supersymmetry 
in two-dimensions and have a BPS domain wall solution. 
It would be interesting to examine this solution in terms of brane 
configurations in ten-dimensions.

%%%%%%%%%%%%%%%%%%%%%%%%%%%%%%%%%%%%%%%%%%%%%%%%%%%%%%%%%%%%%%%%%%
\subsection*{Acknowledgements}
Authors would like to thank K.~Hashimoto, S.~Iso and H.~Nakajima for useful discussions and
 conversations. 
The work of M.~A. is supported by the Science Research Center Program of
 the Korea Science and Engineering Foundation through the Center for
 Quantum Spacetime (CQUeST) of Sogang University with grant number
 R11-2005-021.
M.~A. would like to thank the KEK theory group and the High Energy Physics Division of the Department
 of Physics, University of Helsinki, for their hospitality during his visit.
The work of S.~S. is supported by bilateral exchange program between 
Japan Society for the Promotion of Science (JSPS) and the Academy of 
Finland (AF).

\begin{appendix}
\section{BPS matrices}
The explicit form of the BPS matrices was first constructed in 
\cite{GoRoRaVe} to find a vacuum of the mass deformed ABJM model.
The BPS matrices satisfy the following conditions,
\begin{eqnarray}
& & S^A = S^B S^{\dagger}_B S^A - S^A S^{\dagger}_B S^B, \\
& & S^{\dagger}_A = S^{\dagger}_A S^B S^{\dagger}_B 
- S^{\dagger}_B S^B S^{\dagger}_A.
\end{eqnarray}
The explicit form of the matrices satisfying the above conditions 
is given by 
\begin{eqnarray}
& & (S^{\dagger}_1)_{mn} = \sqrt{m -1} \delta_{mn}, \quad 
(S^{\dagger}_2)_{mn} = \sqrt{N-m} \delta_{m+1,n}, \\
& & S^1 S^{\dagger}_1 = \mathrm{diag} (0,1,2, \cdots, N-1) = 
S^{\dagger}_1 S^1, \\
& & S^{\dagger}_2 S^2  = \mathrm{diag} (N-1, N-2, \cdots, 1,0), \\
& & S^2 S^{\dagger}_2 = \mathrm{diag} (0,N-1,N-2, \cdots, 1), \\
& & S^{\dagger}_A S^A = (N-1) \mathbf{1}_{N \times N}.
\end{eqnarray}
Therefore we have
\begin{eqnarray}
\mathrm{Tr} S^A S^{\dagger}_A = \mathrm{Tr} S^{\dagger}_A S^A = N (N-1).
\end{eqnarray}
These matrices satisfy the following relations,
\begin{eqnarray}
& & \epsilon_{CD} S^C S^{\dagger}_A S^D = \epsilon_{AB} S^B, \quad 
 \epsilon^{CD} S^{\dagger}_C S^A S^{\dagger}_D = - \epsilon^{AB} 
S^{\dagger}_B, \\
& & \epsilon_{AC} \epsilon^{CB} = \delta_A {}^B, \quad 
\epsilon^{AC} \epsilon_{CB} = \delta^A {}_B, \quad 
\epsilon^{12} = - \epsilon_{12} = 1.
\end{eqnarray}

\section{$\mathcal{N} = 2$ superfield formulation of ABJM model}
Here we briefly review the $\mathcal{N}=2$ superfield formulation of the 
 mass-deformed ABJM model given in section 2. 
We follow the same notation as in \cite{BeKlKlSm}.
In terms of ${\cal N}=2$ superfields, 
 the mass-deformed ABJM model is described by chiral 
 superfields ${\cal Z}^A, {\cal W}_A,\bar{\cal Z}_A, \bar{\cal W}^{A}(A=1,2)$ 
 and two vector superfields ${\cal V}, \hat{\cal V}$.
The chiral superfields ${\cal Z}^A, {\cal W}_A(A=1,2)$ 
 and $\bar{\cal Z}_A, \bar{\cal W}^{A}$
transform in the bi-fundamental representations 
$(\bf{N},\bar{\bf N})$ and $(\bar{\bf N},{\bf N})$ 
 under $U(N)\times U(N)$, 
respectively.
They are expanded in components
\begin{eqnarray}
 &&{\cal Z}(x_L)=Z(x_L)+\sqrt{2}\theta \zeta(x_L) +\theta^2 F(x_L), \label{comp1}\\
 &&\bar{\cal Z}(x_R)=Z^\dagger(x_R) - \sqrt{2}\bar{\theta}
  \zeta^\dagger(x_R) - \bar{\theta}^2 F^\dagger(x_R),\label{comp2}\\
&&{\cal W}(x_L)=W(x_L)+\sqrt{2}\theta \omega(x_L) +\theta^2 G(x_L),\label{comp3}\\
 &&\bar{\cal W}(x_R)=W^\dagger(x_R) - \sqrt{2}\bar{\theta}
  \omega^\dagger(x_R) - \bar{\theta}^2 G^\dagger(x_R).\label{comp4}
\end{eqnarray}
Here $x_L^\mu=x^\mu+i(\theta\gamma^\mu\bar{\theta})$ and 
 $x_R^\mu=x^\mu - i(\theta\gamma^\mu\bar{\theta})$, where
 $\gamma^\mu_{\alpha\beta}$ are 
the three-dimensional $\gamma$-matrices defined
 by $\gamma^\mu_{\alpha\beta}=(-{\bf 1},-\sigma^3,\sigma^1)$.
The vector superfields ${\cal V}$ and ${\cal \hat{V}}$ 
transform in the adjoint representations $({\bf adj}, 1)$ and $(1, {\bf adj})$ of $U(N)\times
 U(N)$.
The component expansion in the Wess-Zumino gauge is given by
\begin{eqnarray}
 &&{\cal
  V}(x)=2i\theta\bar{\theta}\sigma+2\theta\gamma^\mu\bar{\theta}A_\mu
   +\sqrt{2}i\bar{\theta}^2\theta\chi
   -\sqrt{2}i\theta^2\bar{\theta}\bar{\chi}+\theta^2\bar{\theta}^2
   D,\label{comp5}
\end{eqnarray}
 and similarly for $\hat{\mathcal{V}}$.
The action in terms of $\mathcal{N}=2$ superfield is written by
\begin{eqnarray}
 S=S_{\rm kin}+S_{\rm CS}+S_{\rm sp},
\end{eqnarray}
where
\begin{eqnarray}
&& S_{\rm kin}=\int d^3x d^4\theta {\rm Tr}
 \left(-\bar{\cal Z}_A e^{-{\cal V}}{\cal Z}^A e^{-\hat{\cal V}}
        -\bar{\cal W}^A e^{-\hat{\cal V}}{\cal W}_A e^{{\cal
	V}}\right), \\
&& S_{\rm CS}=-{ik \over 8\pi}\int d^3x d^4\theta \int_0^1 dt
  {\rm Tr}\left({\cal V}\bar{D}^\alpha(e^{t{\cal V}}D_\alpha
	   e^{-t{\cal V}})
   -\hat{\cal V}\bar{D}^\alpha(e^{t\hat{\cal V}}D_\alpha e^{-t\hat{\cal
   V}})\right), \\
&& S_{\rm sp}={8\pi \over k}\left(
 \int d^3 x d^2\theta W({\cal Z},{\cal W})+\int d^3 x d^2\bar{\theta}
 \bar{\cal W}(\bar{\cal Z},\bar{\cal W})
 \right),
\end{eqnarray}
where the superpotential is given by (\ref{superpotential}),
 (\ref{massless_superpot}) and (\ref{mass_superpot}).
Substituting the expressions (\ref{comp1})-(\ref{comp5}) into the
 action and integrating over the Grassmann measure one obtains the off-shell
 component action.
Further elimination of the set of auxiliary fields yields 
 the on-shell component action.
The equations of motion are given by
\begin{eqnarray}
& \displaystyle
 F^A={2\pi \over k}\left(2\epsilon^{AC}\epsilon_{BD}
 W^{\dagger B}Z_C^\dagger W^{\dagger D}-{km \over 2\pi}W^{\dagger
 A}\right),\label{aux1}&\\
& \displaystyle
 G_A={2\pi \over k}\left(-2\epsilon_{AC}\epsilon^{BD}Z_B^\dagger
		     W^{\dagger C}Z^\dagger_D-{km \over 2\pi}Z_A^\dagger
 \right),\label{aux2}& \\
&N^A=\sigma Z^A - Z^A \hat{\sigma}\,,~~~M_A=\hat{\sigma}W_A-W_A\sigma,\label{aux3}&\\
& \sigma^a={2\pi \over k}{\rm Tr}T^a(ZZ^\dagger - W^\dagger W),~~~~
\hat{\sigma}^a={2\pi \over k}{\rm Tr}T^a(Z^\dagger Z- W
 W^\dagger),\label{aux4}&\\
& \chi^a=-{4\pi \over k}{\rm Tr}T^a (Z \zeta^\dagger - \omega^\dagger
 W),~~~~
\hat{\chi}^a=-{4\pi \over k}{\rm Tr}T^a (\zeta^\dagger Z- W \omega^\dagger).\label{aux5}&
\end{eqnarray}
The bosonic on-shell component action is described in
(\ref{boson-action}), (\ref{boson1}), (\ref{boson2}) and (\ref{boson3}).

\section{$\mathcal{N} = 2$ abelian Chern-Simons-Higgs model from ABJM}
In this appendix, we see the full structure of the effective Lagrangian 
on polarized M2-branes. 
In addition to the ansatz for the bosonic parts (\ref{ansatz}), we 
consider the following fermionic ansatz,
\begin{eqnarray}
\zeta^A = - i \omega^{\dagger A} \equiv \psi S^A, 
\label{ferm_ansatz}
\end{eqnarray}
where $\psi_{\alpha}$ is a complex two component spinor. 
Then the effective Lagrangian corresponding to the ansatz (\ref{ansatz}), 
(\ref{ferm_ansatz}) is
\begin{eqnarray}
\mathcal{L}_{\mathrm{eff}} &=& 2 N (N-1) 
\left[
\frac{k}{16 \pi} \epsilon^{\mu \nu \rho} a_{\mu} f_{\nu \rho} 
- |\mathcal{D}_{\mu} f |^2 - i \psi^{\dagger} \gamma^{\mu} 
\mathcal{D}_{\mu} \psi \right. \nonumber \\
& & \qquad \qquad \qquad \left. 
- \frac{16\pi^2}{k^2} |f|^2 
\left(
|f|^2 - \frac{km}{4\pi}
\right)^2 
+ i \left(\frac{4\pi}{k} \right) 
\left(
3 |f|^2 - \frac{km}{4\pi}
\right) \psi^{\dagger} \psi
\right].
\end{eqnarray}
This result precisely gives $\mathcal{N} = 2$ supersymmetric abelian 
Chern-Simons-Higgs model studied in \cite{LeLeWe} after suitable 
rescaling of variables.
The $\mathcal{N} = 2$ supercharge is given by
\begin{eqnarray}
q = \int \! d^2 x \ 
\left[
\gamma^{\mu} \gamma^0 \psi \mathcal{D}_{\mu} \bar{f} 
- \frac{k}{4\pi} \gamma^0 \psi \bar{f} 
\left(
|f|^2 - \frac{km}{4\pi}
\right)
\right].
\end{eqnarray}
This satisfies the relation
\begin{eqnarray}
\{q_{\alpha}, \bar{q}^{\beta} \} 
= - i (\gamma^{\mu})_{\alpha} {}^{\beta} P_{\mu} 
- \delta_{\alpha} {}^{\beta} Z.
\end{eqnarray}
Here $P_{\mu}$ is the momentum operator and $Z$ is the central charge 
given by
\begin{eqnarray}
Z = \int \! d^2 x 
\left[
- \frac{1}{2} \varepsilon^{ij} f_{ij} |f|^2 + \frac{4\pi}{k} i 
\left(
|f|^2 - \frac{km}{4\pi} 
\right)
\left(
\bar{f} \mathcal{D}_0 f - f \mathcal{D}_0 \bar{f} 
- i \bar{\psi} \gamma^0 \psi
\right)
\right],
\end{eqnarray}
where $\varepsilon^{12} = - 1$. 
The central charge corresponding to the BPS configurations in section 5 is 
easily evaluated. The result is 
\begin{eqnarray}
Z = \frac{km}{4\pi} \int \! d^2 x \ B.
\end{eqnarray}
This correctly reproduces the energy bound discussed in section 5.

\end{appendix}

\end{document}